\documentstyle[12pt]{article}
\catcode`\@=11

\def\marginnote#1{}
\newcount\hour
\newcount\minute
\newtoks\amorpm
\hour=\time\divide\hour by60
\minute=\time{\multiply\hour by60 \global\advance\minute by-\hour}
\edef\standardtime{{\ifnum\hour<12 \global\amorpm={am}%
        \else\global\amorpm={pm}\advance\hour by-12 \fi
        \ifnum\hour=0 \hour=12 \fi
        \number\hour:\ifnum\minute<10 0\fi\number\minute\the\amorpm}}
\edef\militarytime{\number\hour:\ifnum\minute<10 0\fi\number\minute}
\def\draftlabel#1{{\@bsphack\if@filesw {\let\thepage\relax
   \xdef\@gtempa{\write\@auxout{\string
      \newlabel{#1}{{\@currentlabel}{\thepage}}}}}\@gtempa
   \if@nobreak \ifvmode\nobreak\fi\fi\fi\@esphack}
        \gdef\@eqnlabel{#1}}
\def\@eqnlabel{}
\def\@vacuum{}
\def\draftmarginnote#1{\marginpar{\raggedright\scriptsize\tt#1}}
\def\draft{\oddsidemargin -.5truein
        \def\@oddfoot{\sl preliminary draft \hfil
        \rm\thepage\hfil\sl\today\quad\militarytime}
        \let\@evenfoot\@oddfoot \overfullrule 3pt
        \let\label=\draftlabel
        \let\marginnote=\draftmarginnote
   \def\@eqnnum{(\theequation)\rlap{\kern\marginparsep\tt\@eqnlabel}%
\global\let\@eqnlabel\@vacuum}  }

\def\preprint{\twocolumn\sloppy\flushbottom\parindent 1em
        \leftmargini 2em\leftmarginv .5em\leftmarginvi .5em
        \oddsidemargin -.5in    \evensidemargin -.5in
        \columnsep 15mm \footheight 0pt
        \textwidth 250mmin      \topmargin  -.4in
        \headheight 12pt \topskip .4in
        \textheight 175mm
        \footskip 0pt
        \def\@oddhead{\thepage\hfil\addtocounter{page}{1}\thepage}
        \let\@evenhead\@oddhead \def\@oddfoot{} \def\@evenfoot{} }

\def\titlepage{\@restonecolfalse\if@twocolumn\@restonecoltrue\onecolumn
     \else \newpage \fi \thispagestyle{empty}\c@page\z@
        \def\thefootnote{\fnsymbol{footnote}} }

\def\endtitlepage{\if@restonecol\twocolumn \else  \fi
        \def\thefootnote{\arabic{footnote}}
        \setcounter{footnote}{0}}  %\c@footnote\z@ }

\catcode`@=12
\relax
\def\beq{\begin{equation}}
\def\eeq{\end{equation}}

\def\NP#1#2#3{Nucl. Phys. \underline{#1} (19#2) #3}
\def\ov{\overline}
\def\PL#1#2#3{Phys. Lett. \underline{#1} (19#2) #3}

\relax
\begin{document}
\topmargin-2.4cm
\begin{titlepage}
\topmargin 0in
\begin{center}
\hfill {NEIP--94--013}\\
\hfill {hep-th/9412112}\\
\hfill  {September 1994}
\end{center}
\vspace{2.3cm}
\begin{center}{\Large\bf
Anomaly Cancellations and String Symmetries \\ in the
Effective Field Theory}
\vskip 1.2cm
{\bf Jean-Pierre Derendinger}
\vskip .1in
Institut de Physique \\
Universit\'e de Neuch\^atel \\
CH--2000 Neuch\^atel, Switzerland
\end{center}
\vspace{2.4cm}
\begin{center}
{\bf Abstract}
\end{center}
\begin{quote}
This contribution briefly describes some developments
of the use of string symmetries and anomaly cancellation
mechanisms to include string loop corrections
in the construction of the low-energy
effective supergravity of superstrings.
\end{quote}
\vspace{1.2cm}
\begin{center}
{\it Presented at the 27th International Conference on High Energy Physics,
Glasgow, July 1994}
\end{center}
\end{titlepage}
\setcounter{footnote}{0}
\setcounter{page}{0}
\newpage
%
% BODY
%
\section{Effective field theory}

The purpose of the effective low-energy field theory is to describe the
dynamics of massless string modes in the low-energy domain where string massive
states have only virtual effects. This energy range can be characterized by an
ultraviolet physical cutoff $M_{uv}$, which is smaller than the lightest
massive mode of the string theory. The effective field theory is
specified by a local lagrangian density, a Wilson effective lagrangian ${\cal
L}_{eff}$.
In string perturbation theory which is not expected to lead to spontaneous
breaking of supersymmetry \footnote{
For a discussion of the status of supersymmetry breaking in superstrings,
see the contribution by D. L\"ust \cite{L}},
the effective low-energy field theory of a superstring will be a $N=1$
supergravity.

Consider a certain amplitude ${\cal A}(p_1, p_2, \ldots)$ for a physical
process involving only massless external string modes, computed in
string perturbation theory:
\beq
\label{ampl1}
{\cal A}(p_1, p_2, \ldots) = \sum_{L\ge0} {\cal A}^{(L)}(p_1, p_2, \ldots),
\eeq
where $L$ is the string loop order.
Its arguments are external momenta, helicity or internal quantum numbers
attached to the external states and also parameters which must be
introduced to define string perturbation theory.
For instance, the presence of massless modes
requires the introduction of an infrared cutoff $\Lambda$ to regulate loops.
This arbitrary scale parameter may be identified with the ultraviolet cutoff
$M_{uv}$ of the effective field theory, but this is not necessary.
In the limit where all energies and momenta in (\ref{ampl1}) are small compared
with $M_{uv}$, it is expected that (\ref{ampl1}) is reproduced by the same
amplitude computed perturbatively in the quantum field theory defined
by the effective lagrangian ${\cal L}_{eff}$. In correspondance with expansion
(\ref{ampl1}), this effective lagrangian will have
a formal expansion in string-loop order:
\beq
\label{Lloop}
{\cal L}_{eff} = \sum_{k\ge0} {\cal L}_{eff}^{(k)}.
\eeq
At string tree-level in (\ref{ampl1}) ($L=0$) and in the low-energy limit,
the amplitude ${\cal A}^{(0)}(p_1, p_2,$ $\ldots)$ can be obtained
using the tree-level effective lagrangian ${\cal L}_{eff}^{(0)}$ in which
the effect of massive string modes is hidden in non-renormalisable
interactions.
At this order, ${\cal A}^{(0)}$ and ${\cal L}^{(0)}_{eff}$
do not depend on $\Lambda$ and the amplitude
${\cal A}^{(0)}(p_1, p_2, \ldots)$ is the sum of all tree
diagrams obtained with ${\cal L}_{eff}^{(0)}$.
The knowledge of string tree amplitudes allows then in principle to construct
${\cal L}_{eff}^{(0)}$.

In the low-energy limit, the string one-loop contribution to
(\ref{ampl1}), ${\cal A}^{(1)}(p_1,p_2,\ldots)$, corresponds
in the effective field theory to two classes of contributions. Firstly,
the sum of the relevant one-loop diagrams obtained using
${\cal L}_{eff}^{(0)}$ only. Secondly, the new interactions described by
${\cal L}_{eff}^{(1)}$, which are formally already "string one-loop",
lead to a number of tree diagrams generated by ${\cal L}_{eff}^{(0)}+
{\cal L}_{eff}^{(1)}$ and containing one vertex present in
${\cal L}_{eff}^{(1)}$. Since ${\cal A}^{(1)}$ will depend on the
infrared cutoff $\Lambda$, the effective lagrangian will also depend
on $\Lambda$ starting with the one-loop term ${\cal L}_{eff}^{(1)}$.

In general, a Feynman diagram of the effective lagrangian (\ref{Lloop})
with $\ell$ loops will have a "string-loop-order" given by adding to $\ell$
the orders of all vertices, as defined by the expansion (\ref{Lloop}).
Summing all diagrams up to "string-loop-order" $L_{max}$ will provide the
low-energy limit of the amplitude (\ref{ampl1}) computed up to $L_{max}$
string loops.

\section{String gauge symmetries and anomalies}

An important help in the construction of the effective field theory
${\cal L}_{eff}$ is provided by string symmetries which leave a physical
amplitude like (\ref{ampl1}) invariant at each order of string perturbation
theory. These symmetries strongly constrain the form of the effective
lagrangian, even if they are not in general symmetries of ${\cal L}_{eff}$,
which is not a physical object. At string tree-level, the invariance of
${\cal A}^{(0)}$ implies the invariance of ${\cal L}_{eff}^{(0)}$.
In general however,
this symmetry of ${\cal L}_{eff}^{(0)}$ can be anomalous: some
one-loop diagrams which contribute to the effective description of
${\cal A}^{(1)}$ do not respect the symmetry. Then, the invariance of
${\cal A}^{(1)}$ imposes that the effective contributions generated by
${\cal L}_{eff}^{(1)}$ cancel the one-loop anomaly and restore the string
symmetry. Since the knowledge of ${\cal L}^{(0)}_{eff}$ is sufficient to
compute the one-loop anomalous diagrams, the requirement of anomaly
cancellation gives a strong constraint on the form of ${\cal L}_{eff}^{(1)}$.
In some cases, the anomaly-cancellation condition is strong enough to determine
completely the one-loop terms ${\cal L}_{eff}^{(1)}$.
This procedure can be in principle pursued order by order, except if
the existence of non-renormalisation theorems (similar to the
Adler-Bardeen theorem) terminates the argument at the one-loop order.

Green and Schwarz \cite{GS}
found the first example of string symmetries realized
in this "anomaly-cancellation mode" in ten-dimensional (heterotic or type I)
superstrings, which possess space-time [the Lorentz group $SO(1,9)$] and gauge
[$E_8\times E_8$ or $SO(32)$] symmetries. Both symmetries are anomalous in
the tree-level effective lagrangian and their restoration requires specific
contributions in ${\cal L}_{eff}^{(1)}$. The argument can be summarized as
follows, considering for simplicity gauge symmetries only.

\noindent 1)
The theory contains massless fermions, described by Majorana-Weyl spinors,
which couple chirally to gauge fields. The effective
tree-level lagrangian generates then chiral gauge anomalies through one-loop
anomalous diagrams with six external gauge fields.
The formal expression of the anomaly factorises for gauge groups $E_8\times
E_8$ and $SO(32)$, a necessary requirement to be able to cancel it.

\noindent 2)
The theory also contains an antisymmetric tensor field
$b_{\mu\nu}=-b_{\nu\mu}$.
In the tree-level effective lagrangian ${\cal L}_{eff}^{(0)}$, this field
appears through its gauge invariant curl
\beq
\label{curl}
H_{\mu\nu\rho} = \partial_{[\mu}b_{\nu\rho]} - {\kappa\over\sqrt 2}
\omega_{\mu\nu\rho},
\eeq
involving the gauge Chern-Simons form $\omega_{\mu\nu\rho}$ suitably
normalised.
The tree-level lagrangian contains a term proportional to
$H_{\mu\nu\rho}H^{\mu\nu\rho}$, and then an interaction of the form
\beq
\label{Btree}
\partial^\mu b^{\nu\rho}\omega_{\mu\nu\rho},
\eeq
which couples $b_{\mu\nu}$ to two gauge fields.

\noindent 3)
The one-loop contribution ${\cal L}_{eff}^{(1)}$ will contain precisely
the terms necessary to cancel the gauge anomaly: a coupling of $b_{\mu\nu}$
with four gauge fields, and a contact interaction involving six gauge
fields. These terms in ${\cal L}_{eff}^{(1)}$ have to be
gauge variant, and their variation is specified by the chiral anomaly
computed using the tree-level lagrangian. The sum ${\cal L}_{eff}^{(0)}
+{\cal L}_{eff}^{(1)}$ is then gauge non invariant.

In four dimensions, only $U(1)$ gauge symmetries can be realised in the
anomaly-cancellation mode: only abelian (or mixed abelian--nonabelian)
chiral anomalies factorise. Gauge anomaly cancellation can then only appear in
string vacua with gauge groups containing at least a $U(1)$ factor. The
cancellation mechanism, described by Dine, Seiberg and Witten \cite{DSW},
is closely
analogous to ten-dimensional case mentioned above. The same coupling
(\ref{Btree}) is present at tree-level. To cancel the gauge anomaly,
${\cal L}_{eff}^{(1)}$ contains in particular a gauge-variant
term of the form
$$
b_{\mu\nu} \partial^\mu A^\nu
$$
($A_\mu$ is the abelian gauge field), and the chiral anomaly generated by the
triangle diagram is cancelled using the exchange of the antisymmetric tensor.
The one-loop contribution ${\cal L}_{eff}^{(1)}$ plays the r\^ole of a
Fayet-Iliopoulos term which gives a mass to the abelian vector multiplet
removing the anomalous $U(1)$ symmetry from the low-energy symmetry content of
the model.

The global supersymmetrization of this anomaly cancellation mechanism
is very simply described using a linear multiplet \cite{FWZ, S},
which contains the
antisymmetric tensor $b_{\mu\nu}$, a real scalar and a Majorana spinor.
The real linear superfield $L$ is defined by the supersymmetric constraints
${\cal DD}L = \ov{\cal DD} L = 0$. A supersymmetric lagrangian generalizing
the expressions (\ref{curl}) and (\ref{Btree}) is
\beq
\label{susy1}
\int d^4\theta\, F(L-\Omega),
\eeq
omitting the explicit dependence on chiral superfields and superpotential
terms. $\Omega$ is the supersymmetric generalisation of the Chern-Simons
form and the superfield $L-\Omega$ contains (\ref{curl}). Gauge invariance
requires
\beq
\label{deltaL}
\delta L =  \delta\Omega.
\eeq
In general, $\Omega$ is a fixed linear combination of the Chern-Simons forms
of all factors of the gauge group, which is supposed to contain an
anomalous $U(1)$ factor [with superfield $\tilde V$]. The chiral anomaly
can be represented by the non-local expression
\beq
\label{Uanom}
c\int d^2\theta\, WW\, {\cal P}_L \tilde V +{\rm h.c.},
\eeq
where ${\cal P}_L$ is the chiral (non-local) projector and $W$ is the chiral
gauge curvature superfield. Since $\delta\tilde V =
\Lambda+\ov\Lambda$, its gauge variation is
$$
c\int d^2\theta\, WW\, \Lambda + {\rm h.c.}
= -{c\over 4} \int d^4\theta\, (\Lambda+\ov\Lambda)\Omega,
$$
using simple identities. The one-loop contribution to the effective lagrangian
which cancels this anomaly is then clearly of the form
\beq
\label{GSterm}
-{c\over4} \int d^4\theta\,  (L-\Omega)\tilde V \,,
\eeq
the introduction of $L$ being necessary to avoid unwanted non-abelian
anomalies. This superfield expression contains the required coupling
proportional to $b_{\mu\nu}\partial^\mu \tilde V^\nu$.
The one-loop corrected effective lagrangien is then
\beq
\label{susy2}
{\cal L}_{eff} = \int d^4\theta\, \left[ F(L-\Omega) -{c\over4}
(L-\Omega)\tilde V \right],
\eeq
omitting superpotential terms. The supergravity generalisation of this
globally supersymmetric lagrangian is, in the superconformal formalism,
\beq
\label{sugra1}
{\cal L}_{eff} = \left[S_0 \ov S_0 F \left({L-\Omega\over S_0\ov S_0}\right)
-{c\over4}(L-\Omega)\tilde V\right]_D,
\eeq
where $S_0$ is the chiral compensating multiplet (this is "old minimal
supergravity") and $[\ldots]_D$ denotes the real vector density formula of
superconformal tensor calculus.

It is well known that the antisymmetric tensor can be transformed into a
pseudoscalar with a duality transformation. Its supersymmetric
version, which will be discussed in the last section,
transforms the linear multiplet into a chiral one.

\section{K\"ahler symmetry}

The coupling of a chiral matter--super-Yang-Mills system to supergravity
is naturally
invariant under K\"ahler transformations. In the superconformal approach,
the lagrangian density is \cite{CFGVP}
\beq
\label{LagrSC}
{\cal L} = \left[S_0\ov S_0 e^{-{\cal K}/3} \right]_D
+[S_0^3\omega  + f WW]_F,
\eeq
where the K\"ahler potential ${\cal K}(\Sigma,\ov \Sigma e^V)$ is a real
function of the chiral multiplets $\Sigma$, and the introduction of the gauge
vector multiplet $V$ ensures gauge invariance of the theory.
$W$ is the gauge curvature chiral multiplet and the function $f$ which appears
in the chiral density $[\ldots]_F$ is a holomorphic fonction of $\Sigma$.
Gauge transformations, with chiral parameter $\Lambda$ act according to
\beq
\label{gauge}
\Sigma \longrightarrow e^{i\Lambda}\Sigma , \qquad
\ov\Sigma \longrightarrow \ov\Sigma e^{-i\ov\Lambda} , \qquad
e^V  \longrightarrow  e^{i\ov\Lambda} e^V e^{-i\Lambda},
\eeq
so that $\ov\Sigma e^V\Sigma$ is gauge invariant.

The theory (\ref{LagrSC}) is invariant under the
K\"ahler transformation
\beq
\label{Kahler}
\left\{ \begin{array}{rcl}
\omega &\longrightarrow& e^{-\varphi(\Sigma)}\omega \\
S_0 &\longrightarrow& e^{\varphi(\Sigma)/3}S_0 \\
{\cal K} &\longrightarrow& {\cal K}+\varphi(\Sigma) + \ov\varphi(\ov\Sigma)
\end{array}\right. ,
\eeq
$V$, $W$ and $f$ being unaffected. This K\"ahler
transformation is a formal symmetry which indicates that the lagrangian
(\ref{LagrSC}) only depends on the function
\beq
\label{G}
{\cal G} = {\cal K} + \log \omega\ov\omega
\eeq
[choose $\varphi = \log\omega$ in (\ref{Kahler})].
Notice that the quantity $\ov S_0 e^{-{\cal K}/3}$,
which appears in the lagrangian (\ref{Kahler}), is analogous to the argument
$\ov \Sigma e^V$ of ${\cal K}$ itself. Also the K\"ahler invariant combination
$S_0\ov S_0 e^{-{\cal K}/3}$ is similar to the gauge invariant
$\ov\Sigma e^V \Sigma$. The function ${\cal K}$ is then a K\"ahler connection
in the same way as $V$ is the gauge connection. ${\cal K}$ is a
composite multiplet which enters algebraically in lagrangian (\ref{Kahler}).
Since the theory  (\ref{Kahler}) is both K\"ahler and gauge invariant, the
composite K\"ahler connection will appear in fermion covariant derivatives,
together with gauge potentials (in $V$) and also sigma-model covariantization
of kinetic terms.

It is then natural to consider potential anomalies of K\"ahler symmetry, or
more generally of sigma-model local symmetries \cite{LCO, DFKZ1}. The
supersymmetric formalism
sketched in the previous section translates directly to these cases.
For instance, a K\"ahler anomaly would correspond to the chiral $F$-density
\beq
\label{Kanom}
c_K \left[ WW{\cal P}_L K \right]_F
\eeq
($c_K$ is a numerical coefficient),
replacing $\tilde V$ in(\ref{Uanom}) by the K\"ahler connection ${\cal K}$.
And the effective lagrangian for the theory with the linear
multiplet including the anomaly-cancelling Green-Schwarz
one-loop term is \cite{DFKZ1, DQQ}:
\beq
\label{Klag}
\begin{array}{rcl}
{\cal L}_{eff} &=& {\cal L}_{eff}^{(0)} + {\cal L}_{eff}^{(1)}, \\ &&\\
{\cal L}_{eff}^{(0)} &=&
\left[ (L-\Omega) F\left( X , \ldots\right) \right]_D, \\
&&\qquad X =
(L-\Omega)  e^{{\cal K}/3}(S_0\ov S_0)^{-1} \\ &&\\
{\cal L}_{eff}^{(1)} &=&
-{1\over 4}c_K \left[ (L-\Omega) {\cal K}  \right]_D,
\end{array}
\eeq
omitting again the superpotential. The tree-level lagrangian
${\cal L}_{eff}^{(0)}$ is written in a K\"ahler invariant form:
the variable $X$ is invariant and the dots denote
a possible dependence on other
invariant functions of the chiral multiplets.

K\"ahler symmetry is a property of supergravity couplings. In the
superconformal approach, it is directly related to the chiral internal
$U(1)$ part of the conformal superalgebra. Its relation with superstrings
has to do with the fact that certain string symmetries act on the massless
fields of the effective theory with K\"ahler transformations.
An example is target-space duality in $(2,2)$ orbifolds or Calabi-Yau
strings. Considering a idealized model with a unique $(1,1)$ modulus $T$,
the K\"ahler connection for this modulus would be
$$
{\cal K}_T = -3\log(T+\ov T).
$$
Target-space duality acts on $T$ according to
$$
T \quad\longrightarrow\quad {aT-ib \over icT+d}, \qquad ad-bc=1.
$$
On the connection,
$$
{\cal K}_T \quad\longrightarrow\quad {\cal K}_T  +3\log|icT+d|^2,
$$
which is a particular K\"ahler transformation. A K\"ahler anomaly would then
also be a target-space duality anomaly, and the fact that target-space duality
is a quantum string symmetry implies that the effective
lagrangian should include
anomaly-cancelling terms of the form introduced in (\ref{Klag}).

The component expansion of the loop-corrected effective lagrangian (\ref{Klag})
contains, besides the anomaly-cancelling terms,
corrections to the gauge kinetic terms
which depend on the K\"ahler connection. A string one-loop computation of gauge
kinetic terms [i.e. a string amplitude (\ref{ampl1}) with two external gauge
fields and an arbitrary number of moduli] is then able to directly establish
the existence of the quantum correction ${\cal L}_{eff}^{(1)}$. This
calculation of threshold corrections, performed in the context of $(2,2)$
symmetric orbifolds \cite{DKL} (see also \cite{AGNT, MS}), has in fact been at
the origin of the use of K\"ahler anomaly cancellation for loop-corrected
effective supergravities, as developed in ref. \cite{DFKZ1}. It should however
be mentioned that the inclusion of the complete set of (untwisted) moduli
present in generic $(2,2)$ symmetric orbifold, as in \cite{DKL} and
\cite{DFKZ1}, leads to a situation more complicated than the simple example
considered here \footnote{
See also \cite{Louis}, \cite{D} and \cite{LI}.
}.

\section{Axion--antisymmetric tensor duality}

We have up to now discussed anomaly cancellation using the linear multiplet.
This is a natural approach since the Chern-Simons form is the crucial object in
this mechanism and gauge invariance of the tree-level effective lagrangian
associates the Chern-Simons form with the antisymmetric tensor, as in
(\ref{curl}). As already mentioned, duality can always be used to transform the
antisymmetric tensor into a pseudoscalar or the linear superfield into a
chiral one. Suppose for instance that we want to apply duality to lagrangian
(\ref{susy1}). This theory is equivalent with
\beq
\label{dual1}
\int d^4\theta\, \left[ F(U) - (S+\ov S)(U+\Omega)\right],
\eeq
where $U$ is a real vector superfield and $S$ is chiral. The equation of
motion for $S$ indicates that $U+\Omega$ is linear, and hence (\ref{dual1})
and (\ref{susy1}) are equivalent. Or solving the equation of motion for $U$,
\beq
\label{eom}
{\partial\over\partial U} F(U) = S+\ov S,
\eeq
allows to express $U$ as a function of $S+\ov S$ and
leads to the equivalent theory
\beq
\label{dual2}
\begin{array}{l}
\displaystyle{\int} d^4\theta\, \left[ F(U) -
(S+\ov S)U\right]_{U=U(S+\ov S)} \\ \hspace{2.3cm}
+{1\over4}\left( \displaystyle{\int} d^2\theta\, SWW +{\rm h.c.}\right) \\
\equiv \displaystyle{\int}d^4\theta\, G(S+\ov S)
+{1\over4}\left( \displaystyle{\int} d^2\theta\, SWW +{\rm h.c.}\right).
\end{array}
\eeq
The comparison of (\ref{susy1}) and (\ref{susy2}) shows immediately that
the addition of the abelian gauge anomaly cancelling term in the one-loop
effective lagrangian is equivalent to the substitution
\beq
\label{subst}
G(S+\ov S) \quad\longrightarrow\quad G(S+\ov S +{c\over4}\tilde V)
\eeq
in eq. (\ref{dual2}), as observed in ref. \cite{DSW}. The case of K\"ahler
anomaly is similar. The chiral theory dual to ${\cal L}_{eff}^{(0)}$
in eq. (\ref{Klag}) is
\beq
\label{Kdual1}
\begin{array}{l}
\left[ U F(X_U,\ldots) - (S+\ov S)U \right]_D + {1\over4}[SWW]_D, \\ \\
\hspace{2.3cm}
X_U= U e^{{\cal K}/3}(S_0\ov S_0)^{-1},
\end{array}
\eeq
where $U$ is the function of $S+\ov S$ such that
$$
{\partial\over\partial U} UF(X_U,\ldots) = S+\ov S.
$$
The loop correction ${\cal L}_{eff}^{(1)}$ in (\ref{Klag})
corresponds then to the substitution
\beq
\label{Ksubst}
S+\ov S \quad\longrightarrow\quad S+\ov S +{1\over4}c_K {\cal K}.
\eeq

Notice however that the duality transformation does not
respect the string loop expansion
of the effective Wilson lagrangian. The expansion (\ref{Lloop}), which has
been applied
in the linear multiplet formalism, is resummed by the duality
transformation of the linear multiplet $L$ into the chiral superfield $S$.
This observation suggests that the formal equivalence of $L$ and $S$ does not
necessarily mean that the choice of formulating the effective theory of
superstrings either with $S$ or with $L$ is indifferent. Expansion
(\ref{Lloop}) could apply to one version of the theory only, which would then
allow for an easier and more natural field theory interpretation of physical
quantities computed in string perturbation theory.

The calculations performed in $(2,2)$ string models suggest that the fields
contained in the linear multiplet are in closer relationship to string
physical parameters. More precisely, a study of the $E_8$ sector of some
$(2,2)$ orbifolds \cite{DFKZ2,DQQ} shows that the scalar component $C$ of the
linear multiplet is directly related to the renormalised, physical $E_8$
gauge coupling constant $g_\Gamma$:
\beq
\label{phyg}
{1\over \kappa^2\langle C\rangle} -{C(E_8)\over 16\pi^2} = {1\over g^2_\Gamma}
\eeq
[$C(E_8)=30$ is the $E_8$ quadratic Casimir]. On the other hand, the scalar
component $s$ of the chiral multiplet $S$ is related to the bare, unphysical
gauge coupling constant appearing in the Wilson effective lagrangian
${\cal L}_{eff}$, in the term
$$
-{1\over4}{1\over g_W^2} F_{\mu\nu}^AF^{A\,\mu\nu}.
$$
Then,
\beq
\label{gbare}
{1\over2}\langle s+\ov s\rangle = {1\over g_W^2}.
\eeq
Duality can then be viewed as a transformation from bare to physical
quantities, or as a renormalisation-group transformation from some
unified string coupling to a low-energy running coupling.

More realistic theories need however to be considered in order to decide of
the choice of the most appropriate set of low-energy fields, which would
provide the most natural interpretation of string perturbative calculations.
This is also needed to decide whether the linear multiplet is of special
interest in the discussion of superstring effective supergravities.

\end{document}